\documentclass[%
 reprint, amsmath,amssymb, aps,]{revtex4-2}

\usepackage{graphicx}% Include figure files
\usepackage{dcolumn}% Align table columns on decimal point
\usepackage{bm}% bold math
\usepackage{leftidx}
\usepackage{txfonts}
\usepackage{color}
\usepackage{floatrow}

\DeclareUnicodeCharacter{2212}{-}
\newcommand{\dd}{\mathnormal{d}}

\newcommand{\old}[1]{}

\begin{document}

\preprint{APS/123-QED}

\title{Quasi-periodic oscillatory motion of particles orbiting a distorted deformed compact object}% CHANGE THE TITLEEEEEEEEEEEEEEEEEEEEEEEEEEEEEEEEEEEE
\author{Shokoufe Faraji}
 \email{shokoufe.faraji@zarm.uni-bremen.de}%Lines break automatically or can be forced with \\
\author{Audrey Trova}%
 \email{audrey.trova@zarm.uni-bremen.de}
\affiliation{%
 University of Bremen, Center of Applied Space Technology and Microgravity (ZARM), 28359 Germany
}%
\date{February 14, 2021}% It is always \today, today,

\begin{abstract}
This work explores the dynamic properties of test particles surrounding a distorted, deformed compact object. The astrophysical motivation was to choose such background, which could constitute a more reasonable model of a real situation that arises in the vicinity of compact objects with the possibility of having parameters as the extra physical degrees of freedom. This can facilitate associating observational data with astrophysical systems. This work's main goal is to study the dynamic regime of motion and quasi-periodic oscillation in this background, depending on different parameters of the system. Also, we exercise the resonant phenomena of the radial and vertical oscillations at their observed quasi-periodic oscillations frequency ratio of $3:2$.
\end{abstract}

\maketitle

\section{Introduction}

Quasi-periodic oscillations (QPOs) of X-ray power spectral density have been observed at low (Hz) and high (kHz) frequencies in some observations, and they were discovered in the eighties \cite{2000ARA&A..38..717V}. They were also detected in several Black Hole candidates \cite{2006csxs.book..157M}. Quasi-period oscillations have also been observed in supermassive black hole light curves \cite{2008Natur.455..369G,2013ApJ...776L..10L,2015MNRAS.449..467A,2020AcASn..61....2Z,2003ApJ...585..665H,2018ApJ...860L..10S}. Also, recently the source went through Burst Alert Telescope (BAT) onboard Swift \cite{2021arXiv210107030T,2021arXiv210107020M}. QPOs, peak features in the X-ray observed from stellar-mass BHs and neutron stars, are likely to arise from quite near the compact object itself and exhibit frequencies that scale inversely with the black hole mass, and allow us to probe and study the nature of accretion in highly curved space-time \cite{2004ragt.meet....1A,2015ApJ...798L...5Z,2009GReGr..41.1305S, 2011ApJ...726...11J,2012JCAP...09..014B,2013CQGra..30d5010A,2013arXiv1312.2228B,2015ApJ...801..115M,2016EL....11630006B}.

One of the first QPO models is the Relativistic Precession Model (RPM), which identifies the twin-peak QPO frequencies with two frequencies, namely the Keplerian and the periastron frequencies. In the past years, the RPM is served to explain the twin-peak QPOs in several LMXBs \cite{1998ApJ...492L..59S,2004ApJ...617L..45B,2004astro.ph..1464A,2012ApJ...760..138T}. However, this model has some difficulties explaining relatively large observed high frequencies QPO amplitudes and inferred existence of preferred orbits. To modify this model, the concept of orbital resonance models was proposed \cite{2001A&A...374L..19A,2001AcPPB..32.3605K}. Regarding this modification, the high frequencies QPOs (two-picks) are considered as the resonances between oscillation modes of the accreted fluid - the well-known ratio $3:2$ epicyclic resonance model - identify the resonant frequencies with frequencies of radial and vertical epicyclic axisymmetric modes of disc oscillations. The correlation is a cost of resonant corrections to these frequencies \cite{2005AN....326..864A,2016MNRAS.457L..19T}. While the secret of this $3:2$ ratio has still not been clearly revealed. Nevertheless, the oscillations occur only in certain states of luminosity, and hardness \cite{2006csxs.book..157M,2012MNRAS.426.1701B,2016AN....337..398M}, and this phenomenon is not universal \cite{2012MNRAS.426.1701B}. For a review of quasi-periodic oscillations observation and theory, see for example \cite{2020arXiv200108758I, refId0}.

In this paper, we are motivated by the partial success of the above models. By referring to these models, we assume the QPOs are caused by the fundamental epicyclic frequencies associated with the orbital motion of the matter in the accretion disk also their combinations. In fact, with the other kind of resonance with ratios given by small integral numbers like $1:1, 2:1,$ one can also explain observed QPOs frequencies with the same $3: 2$ ratio \cite{2001A&A...374L..19A,2004astro.ph..1464A,2004A&A...425.1075A}.

On the one hand, the properties of these fundamental frequencies, as mentioned above, have been extensively studied in the case of particles motion in the different backgrounds \cite{2003GReGr..35.1503B,2005Ap&SS.300..127A,2005A&A...437..775T,2005ChJAS...5..258Q,2005ragt.meet..315T,2014PhRvD..89j4001G,2016PhyS...91g3004C,2018arXiv180206773B,2018EPJC...78..486S,2019PhRvD.100d4001T,2020PhRvD.102f4023D} among many others. On the other hand, still there are open questions in this area.

We present our work in the background of a distorted, deformed compact object which is static and axisymmetric. This background is the simplest generalization of the so-called $\rm q$-metric up to quadrupole moments \cite{farajiarxive2020}. This metric has two parameters, aside from the central object's mass, namely distortion parameter and deformation parameter, which are not independent of each other. Besides, in the absence of one or other, one can recover either the $\rm q$-metric or distorted Schwarzschild metric \cite{1982JMP....23..680G}. From a dynamical point of view, these parameters can be seen as perturbation parameters of the Schwarzschild spacetime. We explain this metric briefly in Section \ref{space1}. In this respect, the first static and axially symmetric solution with arbitrary quadrupole moment are described by \cite{doi:10.1002/andp.19173591804}. Then \cite{osti_4201189} introduced a static solution with arbitrary quadrupole in prolate spheroidal coordinates. Later, an equivalent form of this metric was found by Zipoy and Voorhees \cite{doi:10.1063/1.1705005,PhysRevD.2.2119} known as $\gamma$−metric or $\sigma$-metric, and later on, by representing this metric in terms of a new parameter $\rm q$, is known as $\rm q$-metric \cite{2011IJMPD..20.1779Q}. This area of study has been discussed extensively in the literature and generalized in many respects \cite{1970JMP....11.2580G,doi:10.1063/1.1666501,PhysRevD.39.2904,Manko_1990,PhysRevD.90.024041,QUEVEDO198513}, among many others. 

There are several motives to study the circular motion of particles in this background. In the relativistic astrophysical study, it is assumed that astrophysical compact objects are described by the Schwarzschild or Kerr space-times. However, besides these setups, others can imitate a black hole's properties, such as the electromagnetic signature \cite{PhysRevD.78.024040}. It is also possible that some astrophysical observations may not be fitted within the general theory of relativity by using the Schwarzschild or Kerr metric \cite{2019MNRAS.482...52S,2002A&A...396L..31A}, like as the mentioned ratio of QPOs. Also, the astrophysical systems are not always isolated as they surrounded by different kinds of matter and radiation. In addition, while in the more realistic scenario, the rotation should take into account; however, it has been shown the possibility of observed resonant oscillations directly when they occur in the inner parts of accretion flow, even if the source of radiation is steady and perfectly axisymmetric \cite{2004ApJ...617L..45B}. Another motivation to choose such background is to constitute a reasonable model of a real situation that arises in the vicinity of this compact object, where it is not always isolated, with the possibility of analytic analysis through exercising parameters of the model where can be treated as the degrees of freedom of the system.

In this paper, we also explore the dynamics of test particles around a distorted, deformed compact object. This discussion can approximate a diluted astrophysical plasma's complex dynamics, where they can be located around the system.

%%%%%%%%%%%%%%%%%%%%%%%%%%%%%%%%%%%%%%%%%%%%%%%%%%%%%5

The paper's organization is as follows: Section \ref{space1} presents the background object and a study of the motion of test particles in this background. While Section \ref{epi3} explains epicyclic frequencies and stable circular geodesics. The parametric resonances present in Section \ref{res4}. Finally, the conclusions are summarized in Section \ref{sum5}.

Throughout this work, we use the signature $(-,+,+,+)$ and geometrized unit system $G=1=c$, otherwise specified there. Latin indices run from $1$ to $3$, while Greek ones take values from $0$ to $3$.

\section{Space-time of distorted deformed compact object} \label{space1}

There is a space-time which is the simplest asymptotically flat solution of Einstein equation with quadrupole moment. In this respect, the first static and axially symmetric solution with arbitrary quadrupole moment were described in \cite{doi:10.1002/andp.19173591804}. Then \cite{osti_4201189} introduced a static solution with arbitrary quadrupole in prolate spheroidal coordinates. Later, Zipoy and Voorhees \cite{doi:10.1063/1.1705005,PhysRevD.2.2119} found an equivalent transformation that leads to a simple solution which can be treated analytically and known as $\gamma$−metric or $\sigma$-metric, and later on, with representing it in terms of a new parameter is known as $\rm q$-metric \cite{2011IJMPD..20.1779Q}.
Now in this work, we consider generalized $\rm q$-metric, which has $\rm q$-metric as the seed metric and considers the existence of a static and axially symmetric external distribution of matter in its vicinity. By its construction, this metric is only valid locally \citep{1982JMP....23..680G,Chandrasekhar:579245}.  In \cite{2020arXiv201015723F} the metric and its circular geodesics are studied. The metric has this form %In general, this is obtained by applying successive transformation of the Zipoy-Voorhees and $p=\delta-1$ to the distorted Schwarzschild metric . 

\begin{align}\label{EImetric}
	{\rm d}s^2 &= - \left( \frac{x-1}{x+1} \right)^{(1+{\alpha})} e^{2\hat{\psi}} \dd t^2+ M^2(x^2-1) e^{-2\hat{\psi}} \nonumber\\
	 &\left( \frac{x+1}{x-1} \right)^{(1+{\alpha})}\left[ \left(\frac{x^2-1}{x^2-y^2}\right)^{{\alpha}(2+{\alpha})}e^{2\hat{\gamma}}\right. \nonumber\\
	 &\left. \left( \frac{\dd x^2}{x^2-1}+\frac{\dd y^2}{1-y^2} \right)+(1-y^2) \dd{\phi}^2\right],\
\end{align}
where $t \in (-\infty, +\infty)$, $x \in (1, +\infty)$, $y \in [-1,1]$, and $\phi \in [0, 2\pi)$. Where $M$ is a parameter that can be identified as the mass of the body generating the field, which is expressed in the dimension of length. The function $\hat{\psi}$ plays the role of gravitational potential, and the function $\hat{\gamma}$ is obtained by an integration of the explicit form of the function $\hat{\psi}$. These are given by

\begin{align} \label{1111}
\hat{\psi} & = -\frac{\beta}{2}\left[-3x^2y^2+x^2+y^2-1\right],\\
\hat{\gamma} & = -2x\beta(1-y^2)\nonumber\\
  &+\frac{\beta^2}{4}(x^2-1)(1-y^2)(-9x^2y^2+x^2+y^2-1).\
\end{align}
This metric contains three free parameters, namely the total mass, quadrupole moments $\alpha$ deformation parameter, and $\beta$ distortion parameter, which are taken to be relatively small and connected to the $\rm q$-metric and the presence of external mass distribution, respectively. In the case of $\beta=0$, this turns to the mentioned $\rm q$-metric, and in the case of $\alpha=\beta=0$ Schwarzschild metric is recovered.This metric may associate the observable effects to these parameters as new physical degrees of freedom. Also, as we have some freedom in defining these momentum variables, we look for ones that minimize computational time and numerical errors. Also, the circular geodesics in this background studied in \cite{farajiarxive2020}. 

The relation between the prolate spheroidal coordinates $(t, x, y, \phi)$, and the Schwarzschild coordinates $(t, r, \theta, \phi)$ is given by

\begin{align}\label{transf1}
 x =\frac{r}{M}-1 \,, \quad  y= \cos\theta.\,
\end{align}

In addition, the related kinematic quantities in this work, specific energy, angular momentum and Keplerian orbital frequency, respectively, read as

\begin{align}
    &E= e^{-\hat{\psi}}{\sqrt{\left(\frac{x-1}{x+1}\right)^{1+\alpha}\frac{x-S}{x-2S}}} ,\\
    &L= \pm {e^{-\hat{\psi}}(1+x)\sqrt{\left(\frac{x+1}{x-1}\right)^{\alpha}\frac{S}{x-2S}}} ,\\
    &\Omega= e^{2\hat{\psi}}\left(\frac{x-1}{x+1}\right)^{\alpha}\sqrt{\frac{x-1}{(x+1)^3}\frac{S}{x-S}} .\
\end{align}
Where

\begin{equation} \label{s}
    S:= 1+\alpha+\beta x-\beta x^3.
\end{equation}

%%%%%%%%%%%%%%%%%%%%%%%%%%%%%%%%%%%
\subsection{Dynamic of particle in the equatorial plane}

\begin{figure}
   \centering
   \includegraphics[width=\hsize]{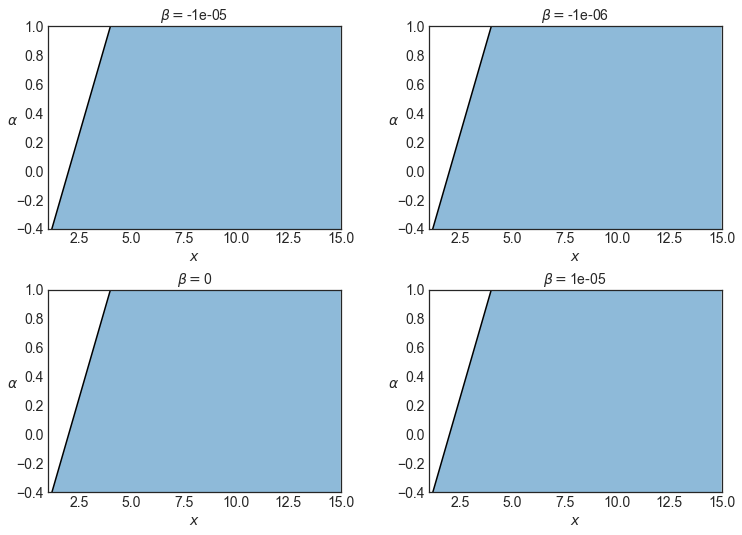}
   \caption{The domain of existence of the circular orbits in the equatorial plane in the ($x,\alpha$)-plane. The lightlike orbits are located on the black curve, $2S-x=0$. The timelike orbits' positions are forming the blue area, which is bound by the locus of the null geodesics $2S-x=0$ and the dashed curve $S=0$.}
    \label{fig1}
\end{figure}

\begin{figure}
  \centering
   \includegraphics[width=\hsize]{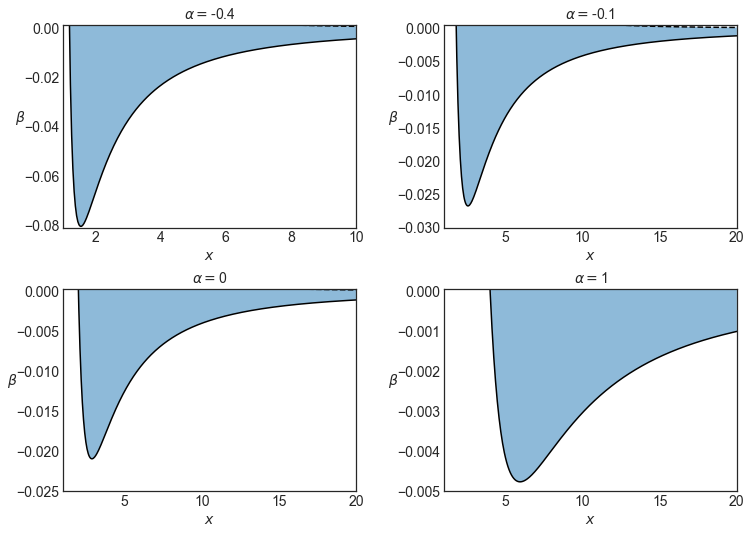}
   \caption{The domain of existence of the circular orbits in the equatorial plane ($x,\beta$) is plane. The lightlike orbits are located on the black curve, $2S-x=0$. The timelike orbits' positions are forming the blue area, which is bound by the locus of the null geodesics $2S-x=0$ and the dashed curve $S=0$.} \label{fig2}
\end{figure}

The effective potential in the equatorial plane is given by this relation \cite{farajiarxive2020}

 \begin{align}\label{Vei}
    V_{\rm Eff}=& \left( \frac{x-1}{x+1} \right)^{(\alpha+1)} e^{2\hat{\psi}}\nonumber\\
    &\left[ \epsilon+\frac{L^2e^{2\hat{\psi}}}{M^2(x+1)^2} \left( \frac{x-1}{x+1} \right)^\alpha\right].\
\end{align}
Regarding, the domain of existence of the circular orbits in the equatorial is plotted in Figures \ref{fig1} and \ref{fig2}. In Figure \ref{fig1} this domain is plotted in the ($x,\alpha$)-plan in terms of the distorted parameter $\beta$, while in Figure \ref{fig2} this region is plotted in the ($x,\beta$)-plane for a range of values of the deformation parameter $\alpha$.

%\begin{figure}
   % \centering
  %  \includegraphics[width=\hsize]{fig4.eps}
   % \caption{Left: Plot of the null-geodesics condition of existence, $2S-x=0$ in the $(x,\alpha)$-plane for different values of $\beta=[-0.005,0.002]$ (upper graph) and in the $(x,\beta)-$plane for different values of $\alpha=[-0.4,5]$ (lower graph). Right : Upper bound of the domain of existence of the timelike orbits, $S=0$, in the $(x,\alpha)-$plane for different values of $\alpha=[-0.005,0.002]$ (upper graph), and in the $(x,\beta)-$plane for different values of $\alpha=[-0.4,5]$ (lower graph).}
 %   \label{fig3}
%\end{figure}
%%%%%%%%%%%%%%%%%%%%%%%%%%%%%%%%%%%%%%%%%%%%%%%%%%%%%%%%%%%%%
%%%%%%%%% TRAJECTORY

\begin{figure*}
    \centering
    %keepaspectratio,height=18cm, width=24cm
    \includegraphics[width=\hsize]{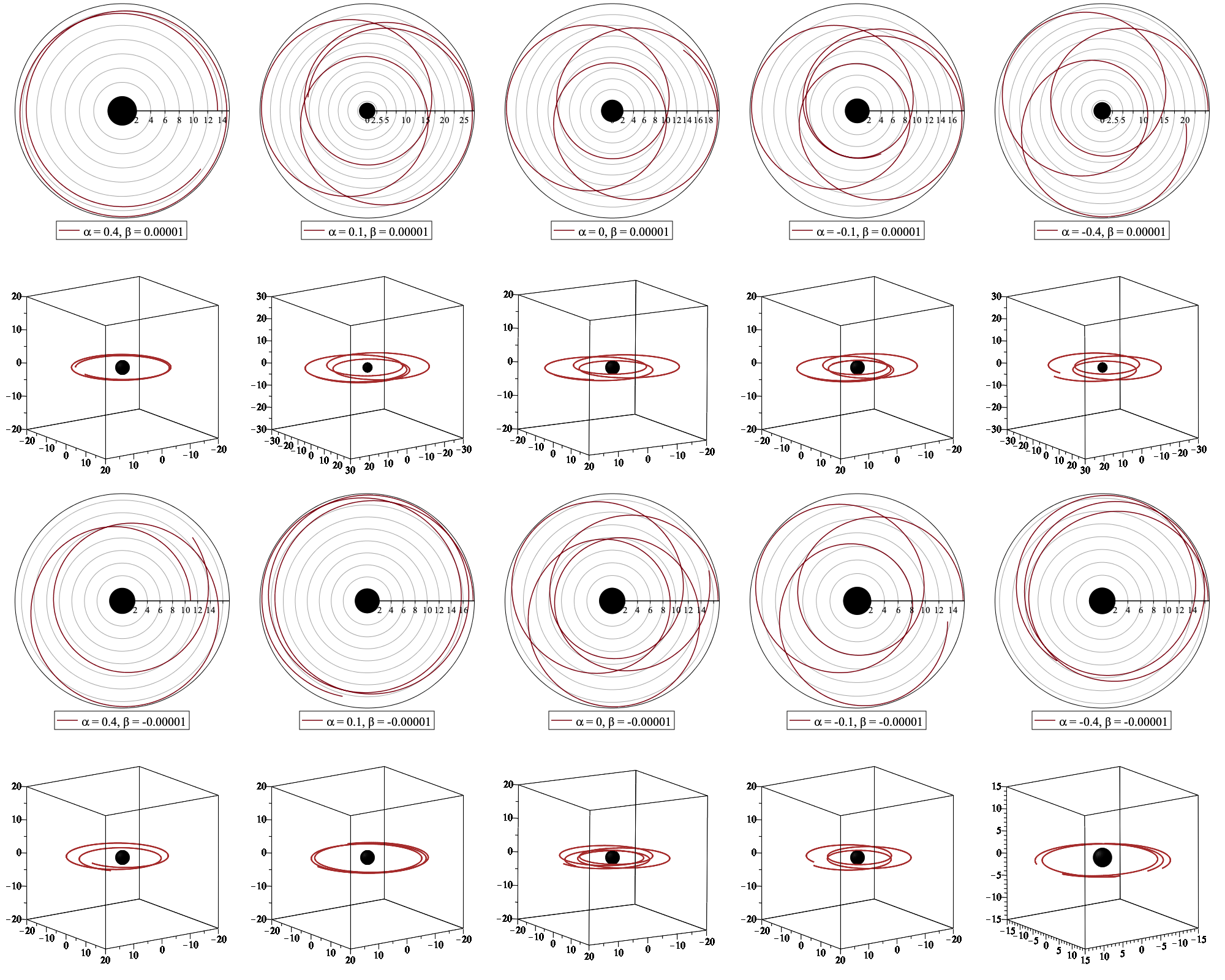}
   \caption{Timelike geodesic for different pairs of $(\alpha,\beta)$. The trajectories in the $(r,\phi)$ section and in the complete 3D are plotted. In the first column, both configurations $\mathcal{E}=0.90$ and $\mathcal{L}=25$. In the second column, both configurations have $\mathcal{E}=0.94$ and $\mathcal{L}=22$. In the central column, both configurations have $\mathcal{E}=0.93$ and $\mathcal{L}=12$. In the fourth column, both configurations have $\mathcal{E}=0.93$ and $\mathcal{L}=15$. In the last one, both configurations have $\mathcal{E}=0.96$ and $\mathcal{L}=9$.}
    \label{bound}
\end{figure*}

The particle motion is limited by the energetic boundaries given by $E^2:=\mathcal{E}= V_{\rm eff}$. By analysing the effective potential, we can have general properties of the dynamics of a particle in this background. In fact, possible types of orbits, in general, dependent on the parameters of $\epsilon$, $\mathcal{E}$, $L$, $\alpha$ and $\beta$. Analytically, dependent on the number of positive real zeros and the sign of $\mathcal{E}-\epsilon$, one obtains different types of trajectories. Here we only discussed bounded timelikes trajectory as we are interested in studying oscillation of particle for a small perturbation of the orbit. In general, test particles' motion can be chaotic in this background for some combinations of parameters $\alpha$ and $\beta$. Here we focus on bounded orbits. Due to the behaviour of the effective potential $V_{\rm eff}$, one can distinguish four different types of the energetic boundary related to having or lacking of initial conditions on both boundaries. Figure \ref{bound} corresponds to the existence of both initial conditions on inner and outer boundaries, where particles trapped in some region forming a toroidal shape around the central object. In this Figure the trajectories for some choices of the parameters are plotted.

%%%%%%%%%%%%%%%%%%%%%%%%%%%%%

%\begin{figure}
   % \centering
    %keepaspectratio,height=18cm, width=24cm
   % \includegraphics[width=\hsize]{fig202.png}
  % \caption{Falling (left) and escape (righ) timelike trajectories for one pair of $(\alpha,\beta)$. The trajectories in the $(r,\phi)$ section and in the complete 3D are plotted. In the first column, the configuration is $\mathcal{E}=0.93$ and $\mathcal{L}=13$. In the second column, $\mathcal{E}=1$ and $\mathcal{L}=22$.}
 %   \label{falsc}
%\end{figure}

%%%%%%%%%%%%%%%%%%%%%%%%%%%%%%%%%%%%%%%%%%%%%%%%%%%%%%%%%%%%%%%
\section{Epicyclic frequencies and stability of circular geodesics}\label{epi3}

In fact, in the accretion disk processes, a variety of oscillatory motions are expected. Indeed, circular and quasi-circular orbits seem to be crucial from the point of view of accretion processes. In the study of the relativistic accretion disk, three frequencies are relevant. The Keplerian orbital frequency $\nu_K = \frac{\Omega}{2\pi}$, radial frequency $\nu_x = \frac{\omega_x}{2\pi}$ and the vertical frequency $\nu_{y} = \frac{\omega_{y}}{2\pi}$. A resonance between these frequencies can be a source of quasi-periodic oscillations that leads to chaotic and quasi-periodic variability in X-ray fluxes observations in many galactic objects.

The Relativistic Precession Model (RPM) is one of the models of study QPO as mentioned earlier. In this model it assumes this is caused by the epicyclic frequencies associated with the quasi-Keplerian motion in the accretion disks. In RPM the upper frequency is defined as the Keplerian frequency $\nu_U = \nu_K$ and the lower frequency is defined as the periastron frequency i.e. $\nu_{p}:= \nu_L = \pm(\nu_K - \nu_x)$. Their correlations are obtained by varying the radius of the associated circular orbit in a reasonable range. Within this framework, it is usually assumed that the variable component of the observed X-ray signal places in a bright localized spot or blob orbiting the compact object on a slightly eccentric orbit. Therefore because of the relativistic effects, the observed radiation is supposed to be periodically modulated.

%%%%%%%%%%%%%%%%%%%%%%%%%%%%%%%%%%%%%%%%%%%%%%%%%%%%

%\subsection{Epicyclic frequencies and stability of circular motion in }

In this section, we explain the stability of circular motion. In this spacetime circular motion in equatorial plane, and the relation between parameters extensively studied in \cite{farajiarxive2020}. 
The equation of motion for a test particle is the geodesic equation

\begin{equation}\label{geo}
    \frac{d^2x^{\mu}}{ds^2}+\Gamma^{\mu}_{\nu\rho} \frac{dx^{\nu}}{ds} \frac{dx^{\rho}}{ds}=0.
\end{equation}
To adapt this equation for this mentioned background, we need to replace all necessary Christoffel symbols \cite{farajiarxive2020}, also substitute $(x=x_0, y=0)$ as we are in the equatorial plane. To describe the more general class of orbits slightly deviated from the circular geodesics in the equatorial plane $x^{\mu}$, we can use the diffeomorphism $x^{\prime \mu}=x^{\mu}+\xi^{\mu}$, and write down geodesic equation \eqref{geo} for this perturbation. By considering terms up to linear order in $\xi^{\mu}$ we obtain \cite{1986Ap&SS.124..137A},

\begin{equation}
   \frac{d^2\xi^{\mu}}{dt^2}+2\gamma^{\mu}_{\ \eta}\frac{d\xi^{\eta}}{dt}+\xi^{\eta}\partial_{\eta}U^{\mu}=0\label{geocom},
 \end{equation}
where 
 
 \begin{align}
  &\gamma^{\mu}_{\ \eta} = \left[2\Gamma ^{\mu}_{\ \eta \delta}u^{\delta}(u^{0})^{-1}\right]_{y=0} ,\\
  &U^{\mu}=\left[\gamma^{\mu}_{\eta}u^{\eta}(u^0)^{-1}\right]_{y=0}.\
\end{align}
Here the 4-velocity for the circular orbits in the equatorial plane is taken as $u^{\mu}=u^0(1, 0, 0, \Omega)$. Then integration of the equation \eqref{geocom} for the $t$ and $\phi$ components leads to

\begin{align}
    &\frac{d\xi^{\eta}}{dt}+\gamma^{\eta}_{\ \nu}\xi^{\nu} = 0,\\
    &\frac{d^2\xi^{x}}{dt^2}+\omega^2_x \xi^{x} = 0 \label{omegay},\\
    &\frac{d^2\xi^{y}}{dt^2}+\omega^2_x \xi^{y} = 0,\
\end{align}
where in the first equation $\eta$ can be taken $t$, or $\phi$; and

\begin{align}
   &\omega^2_{x} = \partial_x U^x - \gamma ^{x}_{\ \eta}\gamma ^{\eta}_{\ x} ,\\
   & \omega^2_{y} = \partial_y U^y .\
\end{align}
This equation system describes the free radial phase and vertical oscillations of a particle around the circular geodesics. For an alternative definition of the epicyclic harmonic motion, see \cite{1984ucp..book.....W}. The sign of frequencies 
$\omega^2_{x}$ and $\omega^2_{y}$ determine the dynamic, so that we have either a stable circular orbits, or even a tiny perturbation can make a strong deviation from the unperturbed path.

%\begin{figure}
 %   \centering
  %  \includegraphics[width=\hsize]{fig3.eps}
   % \caption{The conditions $2S-x=0$ in the $(x,\beta)-$plane has a minimum. We explore the coordinates of this minimum in function of the deformed parameter $\alpha$.}
    %\label{fig3}
%\end{figure}

%  \centering
%\includegraphics[width=\hsize]{fig7.png}
%\caption{Value of the orbital frequency $\nu_K=\frac{c^3}{GM}\frac{\Omega_K}{2 \pi}$ in the plane $(\rm x,\alpha)$ for four values of the distortion parameter $\beta$. $G,c$ are in SI unit and $M=10\textup{M}_{\odot}$. In the white area the orbital frequency is not real.}
%    \label{figkep1}
%\end{figure}

%\begin{figure}[!htbp]
 %   \centering
  %  \includegraphics[width=\hsize]{fig8.png}
   % \caption{Orbital frequency $\nu_K=\frac{c^3}{GM}\frac{\Omega_K}{2 \pi}$ in the plane $(\rm x,\beta)$ for four values of the deformation parameter $\alpha=[-0.6,-0.1,0,1]$. $G,c$ are in SI unit and $M=10\textup{M}_{\odot}$. In the white area the orbital frequency is not real.}
    %\label{figkep2}
%\end{figure}

\begin{figure}
    \centering
    \includegraphics[width=0.6\hsize]{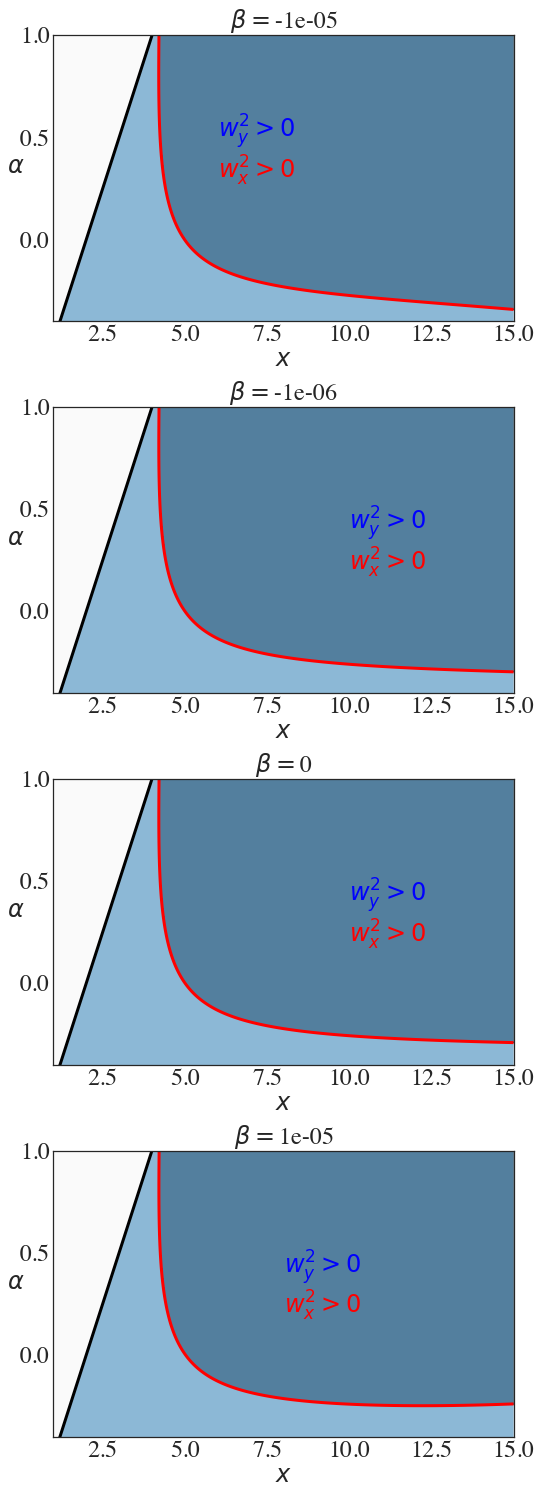}
    \caption{Stability of the timelike circular orbits in the ($x,\alpha$). Timelike circular orbits exist in the light blue area. This region is bounded by the thick dark line $(2S-x=0)$. The red curve represents $w_x^2=0$. The area depicted by the blue-grey region shows the domain of stability. In the chosen range, $w_y^2>0$ in all the light blue region. Thus, note that the timelike orbits are stable above the red curve with respect to vertical perturbations and radial perturbations. On the contrary, below the red curve and in the light blue region, the timelike geodesics are stable with respect to vertical perturbations but unstable with respect to radial perturbations.}
    \label{figst1}
\end{figure}

\begin{figure}
    \centering
    \includegraphics[width=0.65\hsize]{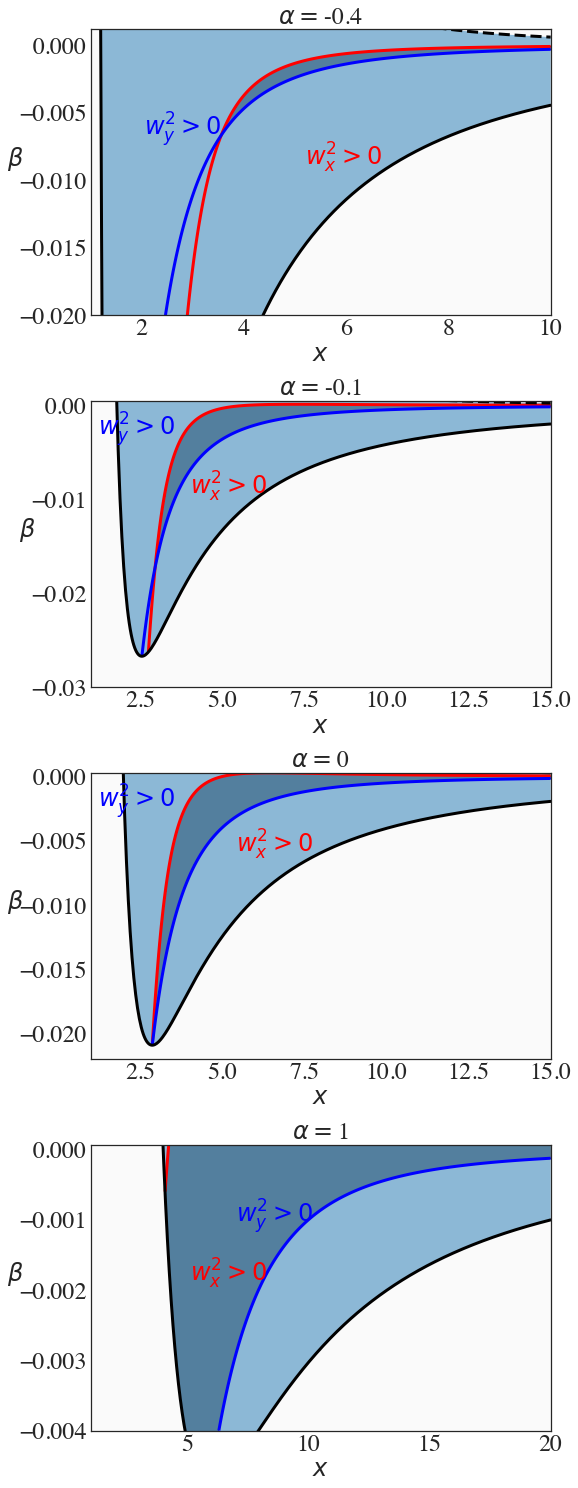}
    \caption{Stability of the timelike circular orbits in the ($x,\beta$)-plane. Timelike circular orbit exists in the light blue area. This region is bounded by the thick dark line $(2S-x=0)$ and the dashed line $(S=0)$. The blue curve represents $w_y^2=0$ and the red one $w_x^2=0$. The blue-grey region shows the stability domain with respect to chosen parameters.}    \label{figst2}
\end{figure}

In the Schwarzschild black ground, these frequencies in spheroidal coordinates are given by

\begin{align} \label{episch}
    &\omega^2_y = \frac{1}{(x+1)^3} ,\\
 &\omega^2_x = \frac{1}{(x+1)^3}\left(1-\frac{6}{x+1}\right).\
\end{align}
The stability of the circular orbits is determined by the radial epicyclic frequency because the vertical frequency coincides with the orbital frequency $\omega^2_y=\Omega$. As it is seen from the equations \eqref{episch} the vertical epicyclic frequency is a monotonically decreasing function of $x$, and we have $\omega^2_x<\omega^2_y=\Omega^2$, also there exists a periapsis shift for bounded quasi-elliptic trajectory implying the effect of relativistic precession that changing the radius of the orbit \cite{2001ASPC..234..213S}. Indeed, this ordering between the frequencies contributes to the possible resonances that may have in a given background. The behaviour of the frequencies helps us to distinguish possible trajectories around a stable circular orbit.

These epicyclic frequencies in the background of a distorted, deformed compact object are written as,

\begin{align}
         w_x^2 = &\frac{\Omega^2 e^{-2\tilde{\gamma}}(1-1/x^2)^{-\alpha(2+\alpha)}}{(x^2-1)} \nonumber \\
     &\left[2(2S-x)(S-x)-\frac{(x^2-1)}{S}(1+\alpha+2\beta x^3)\right],\label{eq:wx2}\\
      w_y^2 = &\frac{\Omega^2 e^{-2\tilde{\gamma}}(1-1/x^2)^{-\alpha(2+\alpha)}}{S}(1+\alpha+2\beta x^3),\label{eq:wy2}\
 \end{align}
where $S$ is given by the relation \eqref{s}. Note that these frequencies are measured concerning the proper time of a comoving observer. The signs of these fundamental frequencies provide a natural condition of having the valid domain of existence of circular and quasi-circular orbits. We explore these frequencies and the valid region more perspicaciously on Figures.% \ref{figkep1}-\ref{figst2}.
%In addition, in the region of existence of the circular orbits the sum of these frequencies is non-negative,

% \begin{equation}
  %   w_y^2 + w_x^2= \frac{\Omega^2 e^{-2\tilde{\gamma}}(1-1/x^2)^{-\alpha(2+\alpha)}}{(x^2-1)}2(2S-x)(S-x).
 %\end{equation}
%{\color{blue} In Figure \ref{figkep1} and \ref{figkep2}, the Keplerian orbital frequency $\nu_K$, in physical unit, for a central source of mass $10(M_\odot)$, is plotted in the plane $(\rm x,\alpha)$ for four values of the distortion parameter $\beta$ and in the $(\rm x,\beta)$ for four values of the deformation parameter $\alpha$. In both figure, in the white area the orbital frequency is not real. We can note that the impact of $\alpha$ and $\beta$ on the Keplerian orbital frequency different. We can see, in the Figure \ref{figkep1}, that the range of frequencies does not evolve a lot in function of $\beta$. On the contrary, in the Figure \ref{figkep2}, increasing $\alpha$ decreases the range of frequencies.}.

In Figure \ref{figst1} and \ref{figst2}, the region of the stability of the timelike circular geodesics is plotted. This region are explored in Figure \ref{figst1} in the $(x,\alpha)$-plane for different values of the distortion parameter $\beta$ and in Figure \ref{figst2} in the $(x,\beta)$-plane for different values of the deformation parameter $\alpha$. In Figure \ref{figst1} as the sign of radial frequency suggests outside of the red curve and in the blue region, the timelike geodesics are unstable for radial perturbations and only are stable for the vertical perturbations. In the Figure \ref{figst2}, both the curves $w_y^2=0$ and $w_x^2=0$ are depicted. The dark region is then bounded by these two curves and show the stable domain. An analysing of these two figures shows the effect of parameter $\alpha$ is more profound rather than $\beta$. In fact, increasing $\beta$ tends to shrink the range of $\alpha$ for allowing to have stability.

The interesting situation in this background, contrary to the Schwarzschild case, is the possibility of various ordering situations that arises among frequencies. For analysing the order of magnitude of these frequencies, we start with,

\begin{equation}
      \frac{w_x^2}{w_y^2}= \frac{2S(2S-x)(S-x)}{(x^2-1)(1+\alpha+2\beta x^3)}-1
 \end{equation}
We analyze this ratio considering the quadrupoles are chosen to be small enough. Also, these parameters $\beta$ and $\alpha$ are not independent of each other; namely, fixing one of them restricts another one's domain. For example, for a small enough parameter $\alpha$, we have this range of orders for parameter $\beta$

\begin{figure*}
    \centering
    \includegraphics[width=\hsize]{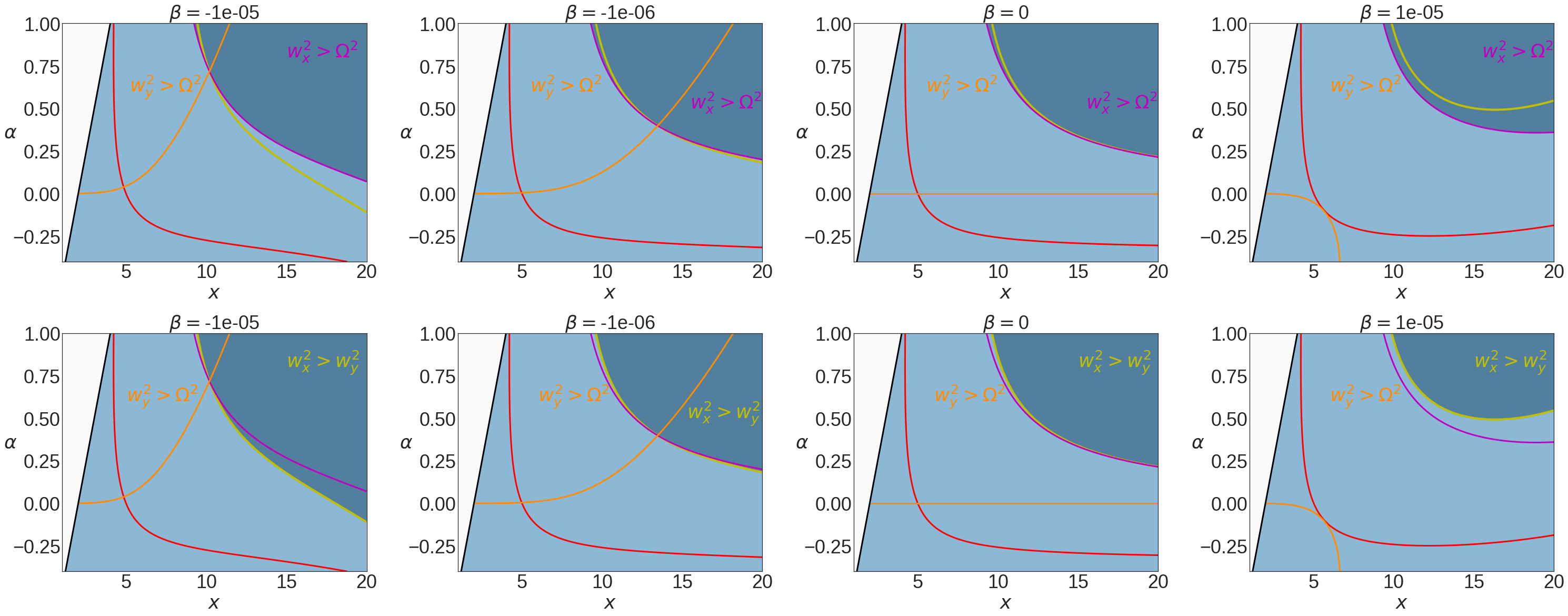}
   \caption{The different epicyclic frequencies in the ($x,\alpha$)-plane are examined. On all the plots, the yellow line depicts $w_x^2=w_y^2$, the pink line shows $w_x^2=\Omega^2$, and the orange line is $w_y^2=\Omega^2$. Moreover, the light blue area corresponds, as in the Figure \ref{figst1} and \ref{figst2}, to the area where circular orbits exist (bounded by the thick black line $2S-x=0$), and above the red line, the orbit is stable with respect to perturbations in both directions. In the panel in the first row, the darker blue area satisfies $w_x^2>\Omega^2$. In the second row, the darker blue area satisfies $w_x^2>w_y^2$. By considering the first and second rows together, one can have different ordering for these frequencies.}
    \label{figorder1}
\end{figure*}

\begin{figure*}
    \centering
    \includegraphics[width=\hsize]{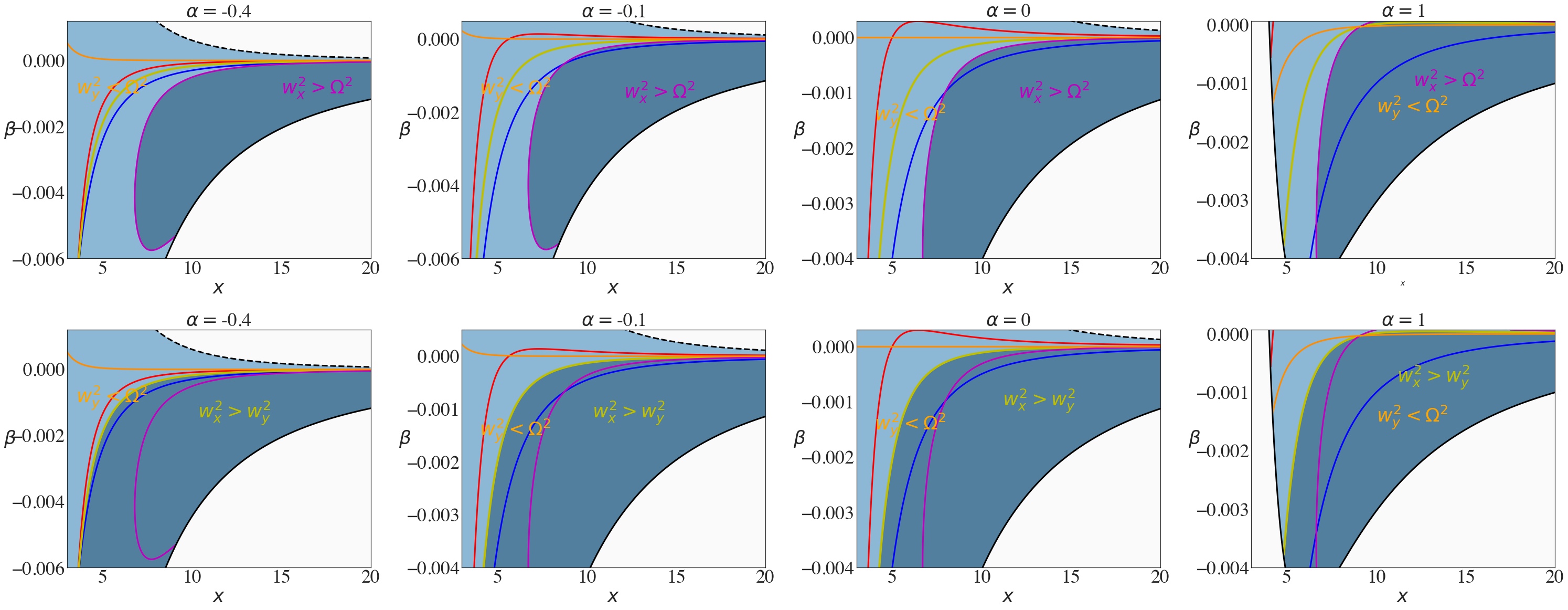}
     \caption{The different epicyclic frequencies in the ($x,\beta$)-plane are examined. As in the Figures \ref{figst1} and \ref{figst2}, the red line shows $w_x^2=0$ and above the red curve both $w_x^2$ and $w_y^2$ are positive. On all the plots, the yellow line depicts $w_x^2=w_y^2$, the pink line shows $w_x^2=\Omega^2$, and the orange line is $w_y^2=\Omega^2$. Furthermore, the light blue area corresponds as in the Figure \ref{figst1} and \ref{figst2} to the area where circular orbits exist (bounded by the thick black line $2S-x=0$ and the dashed line $S=0$). Between the blue and the red line, orbits are stable with respect to the perturbations in both directions. In this panel,  in the first row, the darker blue area satisfies $w_x^2>\Omega^2$. In the second row, the darker blue area satisfies $w_x^2>w_y^2$. By considering the first and second rows together, one can have different ordering for these frequencies.}
    \label{figorder2}
\end{figure*}

\begin{itemize}
  \item $\alpha<0$: \quad $-\mathcal{O}(10^{-2}) < \beta < \mathcal{O}(10^{-3}) - \mathcal{O}(10^{-4})$
  \item $\alpha>0$: \quad $-\mathcal{O}(10^{-3}) < \beta < \mathcal{O}(10^{-4}) - \mathcal{O}(10^{-5})$
\end{itemize}
Using this analysis, it turns out that for a given $\alpha$ and $\beta>0$, we always have $ w_x^2<w_y^2$. However, for $\beta<0$ we need to analyse this condition $S(2S-x)(S-x)$ and regarding its sign we have three different orderings. These various cases are investigated in Figures \ref{figorder1} and \ref{figorder2}, where we compare the three frequencies by plotting the curves of $\omega^2_x$, $\omega^2_y$ and $\Omega^2$. 

In Figure \ref{figorder1} the different epicyclic frequencies in the ($x,\alpha$)-plane for different values and signs of the distortion parameter $\beta$ are plotted. Also, the $w_y^2=\Omega^2$, $w_x^2=\Omega^2$ and $w_x^2=w_y^2$ curves are specified. Moreover, on the left panel, the dark area means $w_x^2>\Omega^2$ and on the right panel $w_x^2>w_y^2$. We illustrated both in one panel to be easy to compare and analyze the behaviour of different regions. In Figure \ref{figorder2} the different epicyclic frequencies are also presented but in the ($x,\beta$)-plane for different values of the deformation parameter $\alpha$. 

By considering the Figures \ref{figorder1} and \ref{figorder2}, we can extract interesting information about the order of the epicyclic frequencies in this background, which strongly influenced by the parameters and their signs and both $\alpha$ and $\beta$ parameters have a crucial role in the order of the frequencies, as we see the lines are intersecting each other multiple times in both planes of discussion.

According to these figures, one can discuss different possibilities. In Figure \ref{figorder1} in the $(x,\alpha)$-plane, if we are interested in the order of the frequencies above the red line where $w_x^2$ and $w_y^2$ are both positives, we see that the ordering is going to be different after intersections. One appears when the red curve crossing the orange curve and the other one when the three curves yellow, pink and orange cross each other, meaning where $\Omega^2=w_y^2=w_x^2$.

Thus, a first possibility appears: Above the red curve, where there is no intersection point, only we have $w_y^2>\Omega^2$. On the contrary, if a crossing point between the red and the orange appears, both orderings $w_y^2>\Omega^2$ and $w_y^2<\Omega^2$ are possible. Following that, a second behaviour appears when $w_y^2>\Omega^2$ and  $w_y^2<\Omega^2$ are possible: appearing the second crossing point makes the order among the three frequencies even more varies. In the following, we give the order for different cases concerning different situations. 
In the last row of Figure \ref{figorder1}, we can extract three different regions:
\begin{itemize}
    \item from the red curve to the pink curve, the order is the following $w_x^2<\Omega^2<w_y^2$.
    \item from the pink line to the yellow line, $\Omega^2<w_x^2<w_y^2$.
    \item from the yellow line to the top of the box, $\Omega^2<w_y^2<w_x^2$.
\end{itemize}
For $\beta \leq 0$, as it has been seen also from Figures Several different regions can appear:
\begin{itemize}
    \item above the red curve and below the orange, three regions are present:
    \begin{itemize}
        \item[-] below the pink and the yellow, we get the following order $w_x^2<w_y^2<\Omega^2$.
        \item[-] above the yellow and below the pink (small area) we get $w_y^2<w_x^2<\Omega^2$.
        \item[-] above the yellow and the pink, $w_y^2<\Omega^2<w_x^2$.
    \end{itemize}
    \item also, above the red curve and above, the orange one, three regions appear:
        \begin{itemize}
        \item[-] below the pink and the yellow, we get the following order $w_x^2<\Omega^2<w_y^2$.
        \item[-] above the pink and below the yellow (small area) we get $\Omega^2<w_x^2<w_y^2$.
        \item[-] above the yellow and the pink, $\Omega^2<w_y^2<w_x^2$.
    \end{itemize}
\end{itemize}
One can produce the same analysis by using Figure \ref{figorder2}. We can note the presence of the two crossing points in the $(x,\beta)-$plane. The analysis of ordering among the different frequencies will be the same as the previous one.

%%%%%%%%%%%%%%%%%%%%%%%%%%%%%%%%%%%%%%%%%%%%%%%%%%%%%%%%%%%%%%%

%\begin{figure}
%    \centering
%    \includegraphics[width=\hsize]{fig6.png}
%    \caption{Orbital frequency $\nu_K=\frac{\Omega}{2 \pi}$ in the plane $(\rm x, %\beta)$ for $\beta=0$ and four values of $\alpha$}
%    \label{mfig6}
%\end{figure}

%%%%%%%%%%%%%%%%%%%%%%%%%%%%%%%%%%%%%%%%%%%%%%%%%%%%%%%%%%%%%%%%
\section{ Parametric resonances}\label{res4}
Before the twin peak, HF QPOs have been discovered in microquasars; this existence and ratio $3: 2$ have pointed out in \cite{2001astro.ph..4487S}. Also, in \cite{2001astro.ph..5057K} authors suggested that the ratios should be rational due to the resonances in quasi-Keplerian accretion disks. Apparently, this fact is well supported by observations. Also, this $3: 2$ ratio as the ratio of $(\nu_{U}:  \nu_{L})$ is seen most often in the twin HF QPOs in the LMXB containing Microquasars. 

In this section we study this phenomena by means of parametric resonance \cite{https://doi.org/10.1002/zamm.19610410910}, and we identify the upper and lower frequencies $(\nu_{U}, \nu_{L})$ with $(\nu_{x}, \nu_{y})$ or $(\nu_{y}, \nu_{x})$. In this respect we explore this fact by standard procedure through the Mathieu's equation. This equation is a linear second-order ODE, which differs from the one corresponding to a harmonic oscillator in the existence of a periodic and sinusoidal forcing of the stiffness coefficient as $f(t)=f_0+f_1\cos(\omega^2_xt)$. The equation then is given by

\begin{equation}
  \frac{d^2\xi^{y}}{dt^2}+\omega^2_y\left[ 1 + \omega^2_yh\cos(\omega^2_xt)\right]\xi^y =0 
\end{equation}
where $h=\frac{f_1}{f_0}\gg 0$ and this a constant, such that the excitation (forcing) term has amplitude $h$, and frequency $\nu_x=\frac{\omega_x}{2\pi}$. The natural, unexcited period is $\nu_y=\frac{\omega_y}{2\pi}$.

It is well known that this set up performs free oscillation around the stable equilibrium case. While, if the stiffness term contains the parametric excitation, i.e. $f_1 \neq 0$, the motion can stay bounded, which is referred to as stable or the motion becomes unbounded, where this case is referred to as unstable (see for example \cite{8f50f04ace1b4bd3a06ea903420c44fc,9147210}).

\begin{figure}
    \centering
    \includegraphics[width=0.62\hsize]{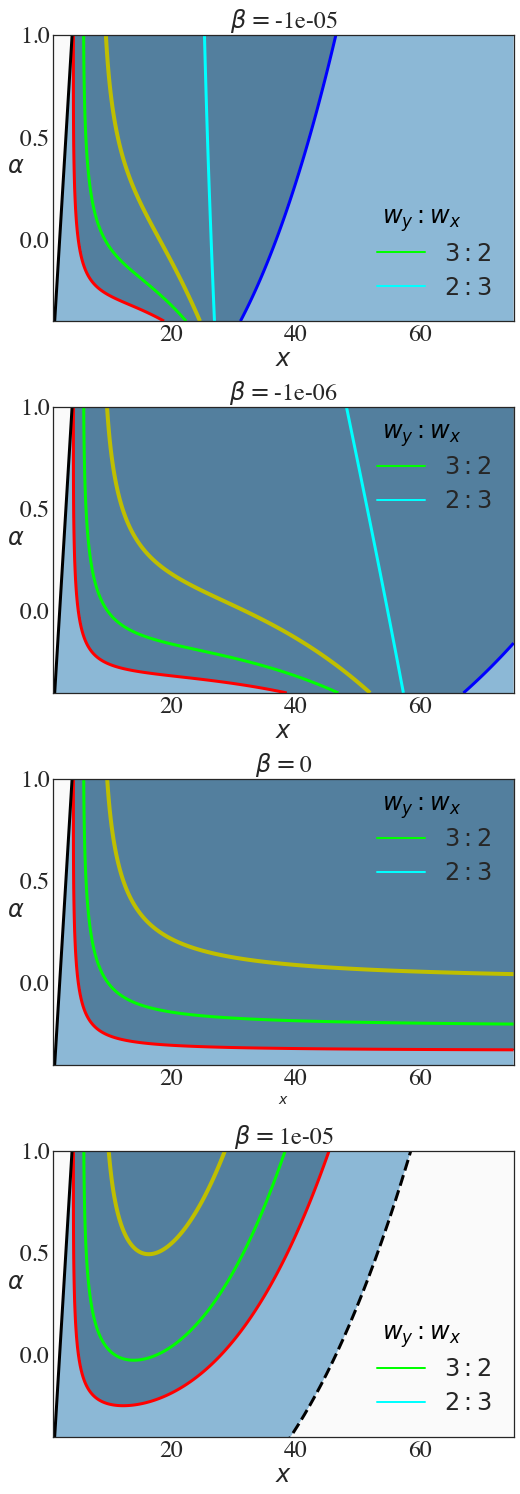}
     \caption{The light blue area depicts the region of existence of the circular orbits. This region is bounded by the thick black curve $(2S-x=0)$ and the dashed line $(S=0)$. The blue-grey region shows the stability of the orbit with respect to vertical and radial perturbations. This area is bounded by the blue curve ($w_y^2=0$) and the red curve ($w_x^2=0$). In this $(x, \alpha)$ plane, the lime curve shows where we have $\frac{w_y}{w_x}=\frac{3}{2}$ and the light cyan curve $\frac{w_y}{w_x}=\frac{2}{3}$.}
    \label{rfig1}
\end{figure}

\begin{figure}
    \centering
    \includegraphics[width=0.7\hsize]{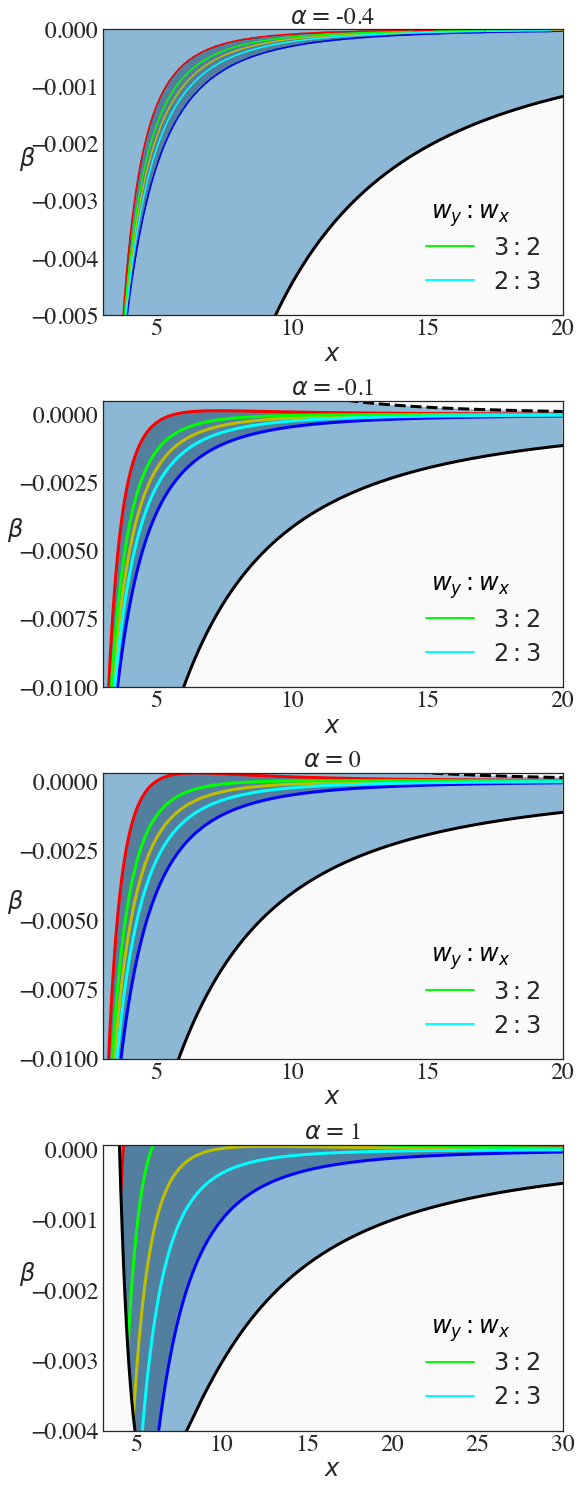}
     \caption{The light blue area depicts the region of existence of the circular orbits. This region is bounded by the thick black curve $(2S-x=0)$ and the dashed line $(S=0)$. The blue-grey region shows the stability of the orbit with respect to vertical and radial perturbations. This region is bounded by the blue ($w_y^2=0$) and the red  ($w_x^2=0$) curves. The lime curve shows the resonance $\frac{w_y}{w_x}=\frac{3}{2}$ and the light cyan curve, shows the resonance $\frac{w_y}{w_x}=\frac{2}{3}$.}
    \label{rfig2}
\end{figure}

The resonance excitation arises for special values of frequencies. In contrast to the standard resonance epicyclic model, the oscillating test particles in this background allow both frequency ratios $\nu_{y} : \nu_{x} = 3 : 2$, and $\nu_{y} : \nu_{x} = 2 : 3$, see Figures \ref{rfig1} and \ref{rfig2}.

Note that in this background, with different combinations of parameters $\alpha$ and $\beta$, also it is possible to have other ratios which can be relevant in other observed data like in other twin frequencies observed in the Microquasar GRS $1915+105$ (see for example \cite{2006astro.ph..7594L}). In Figure \ref{rfig1} and \ref{rfig2}, the resonance ratio $w_x : w_y$ = $3:2$ and $w_x : w_y$ = $2:3$ are depicted. We can see by analysing those two plots that the $3:2$ resonance in the physical range is always present, which is not the case for this ratio $2:3$. For instance, for positive $\beta$, in the possible range of $\alpha$, the $2:3$ resonance ration is not possible.

\begin{figure}
    \centering
    \includegraphics[width=0.75\hsize]{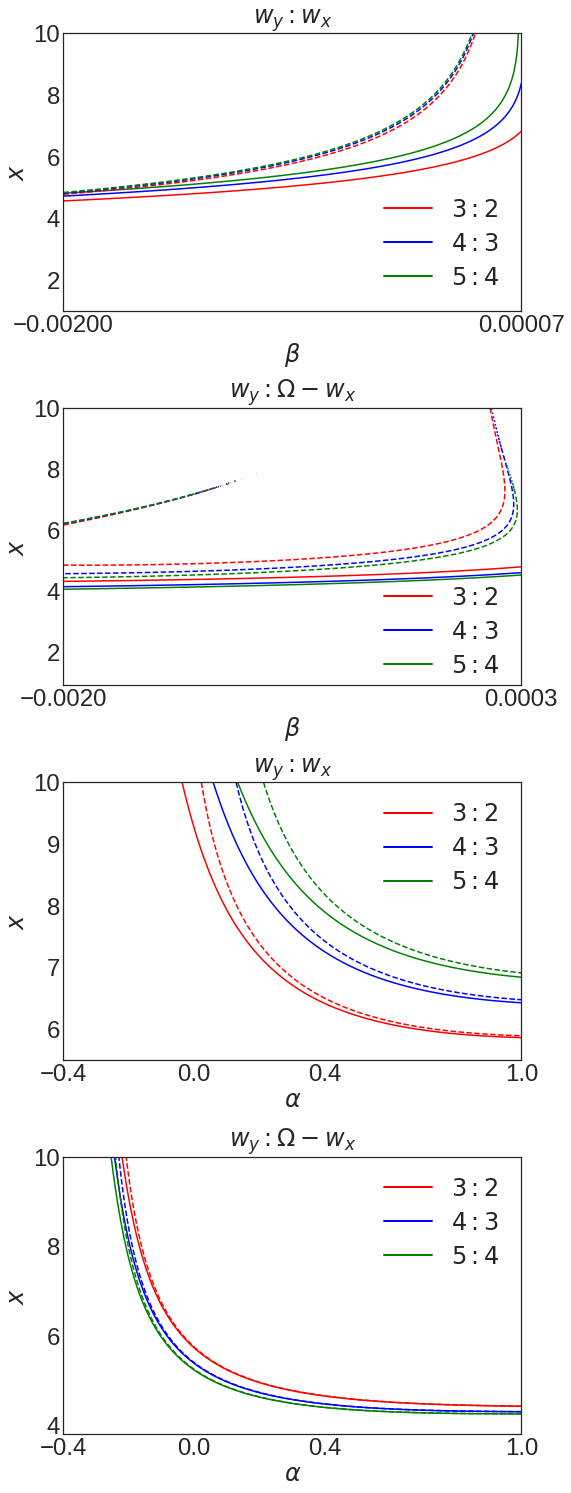}
    \caption{First panel: At the top, the behaviour of the vertical frequency of $w_y$ for three cases of resonances with respect to $\beta$. The thick line shows $\alpha=0.4$ and the dashed line $\alpha=-0.4$. Second panel: location of three epicyclic resonances between vertical epicyclic frequency and periastron frequency in function of $\beta$. The thick line represents $\alpha=0.4$ and the dashed line $\alpha=0$. Third panel: the behaviour of the vertical frequency of $w_y$ for three cases of forced resonances with respect to $\alpha$. The thick line shows $\beta=-0.00001$ and the dashed line $\beta=0.00001$.  Fourth panel: Location of three epicyclic resonances between vertical epicyclic frequency and periastron frequency in function of $\alpha$. The thick line represents $\beta=-0.00001$ and the dashed line $\beta=0$.}
    \label{rfig3}
\end{figure}

In addition, we can identify the frequencies upper and lower frequencies $(\nu_{U}, \nu_{L})$ with different combinations of $\nu_{x}$ and $\nu_{y}$ which can also reproduce the ratio of $3:2$, as mentioned earlier. In Figure \ref{rfig3}, the epicyclic frequencies ratio at
the maximum of $\omega_y$ is depicted. We see that the curves take their minimum at different radius depending on the choice of $\beta$ where they are plotted with respect to $\alpha$. We see that in all cases, almost the positive and negative values of $\beta$ take their maximum at the same radios; however, a further analysis reveal that this radio is smaller for negative values of this parameter. Also, the maximums in all parameters depend on the ratio; for example, we see that as this ratio becomes larger, the maximum happens in smaller radios. This means that the resonance is not monotonic after some distance from the central object, which depends on the combination of parameters in this background.

Of course, considering rotation can modify the radial profiles of the vertical and radial frequencies, which is the subject of future works.

%%%%   DATA DATA    DATA   DATA : : :

%The procedure of fitting the charged particle oscillation frequencies to the observed frequencies is presented in Fig. 13 , for all the three Microquasars GRS $1915+105$, XTE $1550-564$, and GRO $1655-40$. From the restrictions on the spacetime mass parameter M for each of the sources, see Tab. I., we obtain simultaneously restrictions on the external uniform magnetic field through restrictions on the magnetic parameter B. 


%%%%%%%%%%%%%%%%%%%%%%%%%%%%%%%%%%%%%%%%%

%\begin{figure}
 %   \centering
 %   \includegraphics[width=\hsize]{fig22.eps}
 %   \caption{Representation of the upper oscillation frequency $\nu_U$ at the $3:2$ resonance radius. We compared the frequency to the mass-limits of microquasars obtained by observations. Their characteristics are given by the Table \ref{tab:Quasars}.}
  %  \label{figObs}
%\end{figure}

%<<<<<<<<<<<<<<<<<<<<<<<<<<<<<<<<<<<<<<<<<<<<<<<<<<<

%$$$$$$$$$$$$$$$$$$$$$$$$$$$$$$$$$$$$$$$$$$$$$$$$$$$$$

\section{Summary and conclusion}\label{sum5}
In this work, we studied the dynamics of particles and the quasi-periodic oscillation by studying fundamental frequencies of the circular motion around a deformed compact object up to the quadrupole. This background has $\rm q$-metric as the seed metric while considering a distribution of matter in its vicinity. This metric is static and axisymmetric up to quadrupoles also in the external surrounding matter and contains two parameters: distortion parameter $\beta$ and deformation parameter $\alpha$. This background was briefly explained in Section \ref{space1}. These two parameters' dependency reflects into motion and epicyclic frequencies of particles that cost strong deviation from the correspondence quantities in the Schwarzschild case. In this respect, one can explore different orderings among fundamental frequencies and various possibilities to reproduce the ratio of $3:2$ via different combinations of parameters which is not the case in either Schwarzschild or in $\rm q$-metric.

Interestingly, also it is possible from observational data to assign some restrictions on the parameters in this metric. In this regard, also the construction of a thin accretion disk can limit having some combinations of these parameters and modify the metric in this sense \cite{2020arXiv201106634F}.

A further step of this work can be considering rotation that leads to modifying the radial profiles of the vertical and radial frequencies, which is the subject of our future work. Also, the magnetic field can serve as a fundamental input in this system to model more real astrophysical systems. One can also extend this work from a single particle to a complex system like accretion disks. It is also of some interest to apply these models as the initial conditions in the numerical simulations and test their ability to account for observable constraints of astrophysical systems.

%%%%%%%%%%%%%%%%%%%%%%%%%%%%%

%\begin{table}[h!]
%\label{tab:Quasars}
%\centering
 %\begin{tabular}{c c c c} 
 %\hline
% Source &  GRO 1655-40 & XTE 1550-564 &  GRS 1915+105 \\ 
 %\hline\hline
%  $\nu_U$ & 447 — 453 & 273 — 279 &  165 — 171 \\ 
% $\nu_L$ & 295 — 305 & 179 — 189 & 108 — 118 \\
% $\frac{M}{{M}_{\odot}}$ & 6.03 — 6.57 & 8.5 — 9.7 & 9.6 — 18.4 \\
% $a$ & 0.65 — 0.75 &  0.29 — 0.52 &  0.98 — 1 \\[1ex]
 
 % \hline
% \end{tabular}
%\end{table}

\section*{Acknowledgements}
The authors thank the research training group GRK 1620,” Models of Gravity”, funded by the German Research Foundation (DFG).

\bibliographystyle{unsrt}
\bibliography{bibunmagpq}

\end{document}